\UseRawInputEncoding
\documentclass[prd,aps, preprintnumbers, showpacs,twocolumn, nofootinbib,superscriptaddress,notitlepage]{revtex4-1}
\usepackage[normalem]{ulem}
\usepackage{epsfig}
\usepackage{amsfonts}
\usepackage{amsmath}
\usepackage{slashed}
\usepackage{graphicx}
\usepackage{color}
\usepackage{mathtools}
\usepackage{subfigure}
\usepackage{graphicx}

\usepackage{stmaryrd}
\usepackage{amssymb,psfrag}
\usepackage[normalem]{ulem}
\usepackage{caption}
\usepackage{bm}

\newcommand{\beq}{\begin{eqnarray}}
\newcommand{\eeq}{\end{eqnarray}}

\newcommand{\nn}{\nonumber}

\def\red{\textcolor[rgb]{1.00,0.00,0.00}}

\begin{document}

\title{The production of charmonium pentaquark from b-baryon and B-meson decay: SU(3) analysis}

\author{Wei-Hao Han}
\email{hanweihao2022zzu@163.com}
\affiliation{School of Physics and Microelectronics, Zhengzhou University, Zhengzhou, Henan 450001, China}

\author{Ye Xing}
\email{Corresponding author. xingye\_guang@cumt.edu.cn}
\affiliation{School of Physics, China University of Mining and Technology, Xuzhou 221000, China}

\author{Ji Xu}
\email{xuji\_phy@zzu.edu.cn}
\affiliation{School of Physics and Microelectronics, Zhengzhou University, Zhengzhou, Henan 450001, China}

%\date{\today}
\begin{abstract}
In this paper, we study the production of charmonium pentaquark $c \bar c q q q$ from bottom baryon and B-meson decays under the flavor SU(3) symmetry. Decay amplitudes for various processes are parametrized in terms of the SU(3) irreducible nonperturbative amplitudes. A number of relations between decay widths have been deduced. Moreover, the strong decays of pentaquark is also taken into account. These results can be tested in future measurements at LHCb, Belle II and CEPC. Once a few decay branching fractions have been measured, our work could provide hints for exploring new decay channels or new pentaquark states.
\end{abstract}

\maketitle

\section{Introduction}
In 2015, the observation of $J/\psi \, p$ resonances consistent with charmonium pentaquark states in $\Lambda_b^0 \to J/\psi K^- p$ decays were reported by the LHCb Collaboration \cite{LHCb:2015yax}. In practice, states that decay into $J/\psi \, p$ may have distinctive signatures \cite{Li:2014gra}, the minimal quark content can be identified as $c \bar c u u d$, and thus is charmonium pentaquark. Although the existence of pentaquark, which composed of four quarks and an antiquark, has been predicted since the establishment of quark model \cite{Gell-Mann:1964ewy,Zweig_CERN-TH-401CERN-TH-401,Lipkin:1987sk}, the experimental search has taken a pretty long time. Such new particles dramatically changed our understanding of exotic states which can not be included in the conventional quark-antiquark and three-quark
schemes of standard spectroscopy. These charmonium pentaquarks are labeled as $P_c$ which carries an electric charge and couples to charmonium. In addition, they are the first exotic states observed in the heavy-flavor baryonic sector.

Subsequently, a series of pentaquark candidates were reported. In 2019, the LHCb Collaboration updated their analysis of $\Lambda_b^0 \to J/\psi K^- p$ and found a new state $P_c(4312)$ \cite{LHCb:2019kea}. In 2020, a new structure in the $J/\psi \Lambda$ invariant mass distribution, consistent with a charmonium-like pentaquark with strangeness $P_{cs}(4459)$, was obtained from an amplitude analysis of $\Xi_b^- \to J/\psi \Lambda K^-$ decays \cite{LHCb:2020jpq}. In 2022, evidence for a charmonium pentaquark $P_c(4337)$ in the $J/\psi \, p$ and $J/\psi \, \bar p$ systems was found in $B_s^0 \to J/\psi \, p \, \bar p$ decays \cite{LHCb:2021chn}. In 2023, an amplitude analysis of $B^- \to J/\psi \Lambda \, \overline{p}$ is performed, a narrow resonance in the $J/\psi \Lambda$ system, consistent with a pentaquark candidate with strangeness is observed \cite{LHCb:2022ogu}. It seems that we will experience a new era with more and more such exotic states observed in the near future, therefore it is of prime importance to understand the sub-structure of these pentaquarks as well as provide information for exploring new pentaquark states on experiment.

These experimental progresses have made a great impact on the hadron spectroscopy and evoked a lot of theoretical interest. Proposed interpretations of pentaquarks include the compact pentaquark scenarios \cite{Santopinto:2016pkp,Deng:2016rus,Maiani:2018tfe,Wang:2015epa,Zhu:2015bba}, the molecular models \cite{Du:2019pij,Wang:2019ato,Chen:2015loa,Chen:2015moa,Liu:2019tjn}, the hadrocharmonium model \cite{Eides:2019tgv,Kubarovsky:2015aaa}, or peaks due to triangle-diagram processes \cite{Liu:2015fea,Meissner:2015mza,Guo:2015umn}. Besides, there are also studies on the properties of other pentaquark candidates with different quark components \cite{Chen:2015sxa,Feijoo:2015kts,Cao:2019gqo,Xing:2022aij,Li:2023kcl}. Despite of these encouraging results in literatures, we should stress here that the precise structures of pentaquarks remain unknown; there is no consensus as to the explanation of how the five quarks, i.e., the four quarks and an antiquark, are dynamically structured. At this moment, both the experimental and theoretical studies are not yet conclusive. It is widely recognized that to disentangle the various models and further understand the nature of charmonium pentaquark, searches for additional productions and decay channels are crucial \cite{Olsen:2017bmm}.

The decays of $B_s^0$ offer us a cleaner environment to search for pentaquark in stead of baryonic $\Lambda_b^0$ and $\Xi_b^-$ decays \cite{LHCb:2021chn,Xing:2022uqu}. However, whether through baryonic decay or mesonic decay, calculating the decay amplitudes of these transitions is a formidable challenge, there is no factorization approach established to handle production processes of $P_c$. On the other hand, flavor SU(3) symmetry can be used to relate various relevant decays and provide very useful guide for the future pentaquark searches. One significant advantage of the SU(3) analysis is that it is independent of the factorization details, allowing us to relate various decay modes despite the unknown nonperturbative dynamics of QCD \cite{Savage:1989ub,Gronau:1994rj,He:1998rq,Deshpande:1994ii,Chiang:2003pm,
Li:2007bh,Zhou:2015jba,Jiang:2017zwr,Wang:2008rk,Cheng:2014rfa,He:2014xha,Lu:2016ogy,Cheng:2012xb,Li:2012cfa,Qin:2013tje,
Wang:2017azm,Shi:2017dto,Wang:2018utj,Li:2021rfj,Zhang:2018llc,Huang:2021jxt,Huang:2022zsy,He:2006ud}. Certain theoretical models predict that some of these charmonium pentaquarks belong to octet multiplet of flavor SU(3) \cite{Li:2015gta,Ali:2016dkf}, thus finding the other states in the multiplet will provide key evidence for these models.

In this work, we consider the production of charmonium pentaquark from b-baryon and B-meson decays by utilizing flavor SU(3) analysis. Some testable relations for b-baryon decays into a pentaquark plus a light meson and B-meson decays into a pentaquark plus a light baryon are presented. Afterwards, the strong decays of charmonium pentaquark will be discussed as well. Some particular processes can be used as signatures to reconstruct pentaquark. The main motivation of this work is to provide suggestions which may help experimentalists find new $P_c$ states or new production and decay modes of already observed $P_c$.

The rest of this paper is organized as follows. In Sec.\,\ref{mutipltes}, we will collect the the irreducible forms for the particle multiplets in the SU(3) symmetry. In Sec.\,\ref{Production}, we will analyze the nonleptonic decays of b-baryon and B-meson. The strong decays of charmonium pentaquark are investigated in Sec.\,\ref{Strongdecay}. Finally, we conclude in Sec.\,\ref{conclusions}.

%%%%%%%%%%%%%%%%%%%
\section{Particle mutipltes}
\label{mutipltes}
%%%%%%%%%%%%%%%%%%%
In this section, we will collect the representations for the hadron multiplets involved in our work. Under the flavor SU(3) symmetry, the $b$ quark is a singlet and the light quark $q$ belongs to the fundamental representation $3$. Thus the b-baryon contains an antitriplet and a sextet in the SU(3) space which are denoted as $\mathcal{B}$ and $\mathcal{C}$
\begin{eqnarray}
&& \left(\mathcal{B}\right)^{i j} = \left(\begin{array}{ccc}
0 & \Lambda_{b}^{0} & \Xi_{b}^{0} \\
-\Lambda_{b}^{0} & 0 & \Xi_{b}^{-} \\
-\Xi_{b}^{0} & -\Xi_{b}^{-} & 0
\end{array}\right)\,, \nn\\
&& \left(\mathcal{C}\right)^{i j} = \left(\begin{array}{ccc}
\Sigma_{b}^{+} & \frac{\Sigma_{b}^{0}}{\sqrt{2}} & \frac{\Xi_{b}^{\prime 0}}{\sqrt{2}} \\
\frac{\Sigma_{b}^{0}}{\sqrt{2}} & \Sigma_{b}^{-} & \frac{\Xi_{b}^{\prime-}}{\sqrt{2}} \\
\frac{\Xi_{b}^{\prime 0}}{\sqrt{2}} & \frac{\Xi_{b}^{\prime-}}{\sqrt{2}} & \Omega_{b}^{-}
\end{array}\right) \,.
\end{eqnarray}
The bottom meson forms an SU(3) antitriplet:
\begin{eqnarray}
  B_i = \left(\begin{array}{ccc} B^-, & \overline{B}^0, & \overline{B}^0_s \end{array} \right) \,.
\end{eqnarray}
The charmonium pentaquark discussed in this work contains at least three light quarks in addition to a $c\bar c$ pair, i.e. [$c\bar c q q q$].
Under the flavor SU(3) symmetry, the heavy quarks are singlet, the light quark transforms under the flavor SU(3) symmetry as $3 \otimes 3 \otimes 3 = 1\oplus8\oplus8\oplus10$. We denote the octet pentaquark as
\begin{eqnarray}\label{multiPc}
  \mathcal{P}_i^j = \left(\begin{array}{ccc}
\frac{P_{\Sigma^0}}{\sqrt{2}}+\frac{P_{\Lambda}}{\sqrt{6}} & P_{\Sigma^{+}} & P_p \\
P_{\Sigma^{-}} & -\frac{P_{\Sigma^0}}{\sqrt{2}}+\frac{P_{\Lambda}}{\sqrt{6}} & P_n \\
P_{\Xi^{-}} & P_{\Xi^0} & -\frac{P_{\Lambda}}{\sqrt{6}}
\end{array}\right) \,.
\end{eqnarray}
Discoverying these pentaquarks in the multiplet is one of the way to verify the relevant theoretical model.

For the meson sector, the light pseudoscalar mesons form an octet:
\begin{eqnarray}
 (M_{8})_i^j=\begin{pmatrix}
 \frac{\pi^0}{\sqrt{2}}+\frac{\eta}{\sqrt{6}}  &\pi^+ & K^+\\
 \pi^-&-\frac{\pi^0}{\sqrt{2}}+\frac{\eta}{\sqrt{6}}&{K^0}\\
 K^-&\overline{K}^0 &-2\frac{\eta}{\sqrt{6}}
 \end{pmatrix} \,.
\end{eqnarray}
Here $\eta$ is only considered as a member of octet, while the singlet $\eta_1$ is not considered to avoid the octet-singlet mixture complexity.

Light baryons made of three light quarks are presented as:
\begin{eqnarray}
  T_{8}=\left(\begin{array}{ccc}
\frac{1}{\sqrt{2}} \Sigma^{0}+\frac{1}{\sqrt{6}} \Lambda^{0} & \Sigma^{+} & p \\
\Sigma^{-} & -\frac{1}{\sqrt{2}} \Sigma^{0}+\frac{1}{\sqrt{6}} \Lambda^{0} & n \\
\Xi^{-} & \Xi^{0} & -\sqrt{\frac{2}{3}} \Lambda^{0}
\end{array}\right) \,. \nn\\
\end{eqnarray}

The singly charmed baryons can form an antitriplet or sextet. In former case, we have the matrix expression:
\begin{eqnarray}
 T_{\bf{c\bar 3}} &=& \left(\begin{array}{ccc} 0 & \Lambda_c^+  &  \Xi_c^+  \\ -\Lambda_c^+ & 0 & \Xi_c^0 \\ -\Xi_c^+   &  -\Xi_c^0  & 0
  \end{array} \right)\,,
\end{eqnarray}
and in latter case:
\begin{eqnarray}
 T_{\bf{c 6}} &=& \left(\begin{array}{ccc}
\Sigma_{c}^{++} & \frac{1}{\sqrt{2}} \Sigma_{c}^{+} & \frac{1}{\sqrt{2}} \Xi_{c}^{\prime+} \\
\frac{1}{\sqrt{2}} \Sigma_{c}^{+} & \Sigma_{c}^{0} & \frac{1}{\sqrt{2}} \Xi_{c}^{\prime 0} \\
\frac{1}{\sqrt{2}} \Xi_{c}^{\prime+} & \frac{1}{\sqrt{2}} \Xi_{c}^{\prime 0} & \Omega_{c}^{0}
\end{array}\right) \,.
\end{eqnarray}

The anticharmed meson forms an SU(3) triplet:
\begin{eqnarray}
\overline{D}^i = \left(\begin{array}{ccc} \overline{D}^0, & D^-, & D^-_s  \end{array} \right) \,.
\end{eqnarray}

Here we also present the best determination of the magnitudes of CKM matrix elements~\cite{Zyla:2020zbs}
\begin{eqnarray}\label{CKMNum}
&& \left[\begin{array}{lll}
\left|V_{u d}\right| & \left|V_{u s}\right| & \left|V_{u b}\right| \\
\left|V_{c d}\right| & \left|V_{c s}\right| & \left|V_{c b}\right| \\
\left|V_{t d}\right| & \left|V_{t s}\right| & \left|V_{t b}\right|
\end{array}\right] = \nn\\
&& \left[\begin{array}{ccc}
0.97370 \pm 0.00014 & 0.2245 \pm 0.0008 & 0.00382 \pm 0.00024 \\
0.221 \pm 0.004 & 0.987 \pm 0.011 & 0.0410 \pm 0.0014 \\
0.0080 \pm 0.0003 & 0.0388 \pm 0.0011 & 1.013 \pm 0.030
\end{array}\right] \,, \nn\\
\end{eqnarray}
for the convenience of the subsequent discussions.

To describe the various decay modes in the frame of SU(3) analysis, we need to construct the hadron-level effective Hamiltonian with the representations for initial and final states listed above. It is worth stressing that a hadron in the final state must be created by its antiparticle field. For instance, we need a $\overline{P}_{\Lambda}$ field in Hamiltonian to create a $P_{\Lambda}$ pentaquark in the final state. The constructions of hadron-level effective Hamiltonian are displayed in the following sections, they will result in strikingly simple relations among the decay amplitudes.

%%%%%%%%%%%%%%%%%%%%%%%
\section{Production of pentaquark from b-baryon and B-meson}
\label{Production}
%%%%%%%%%%%%%%%%%%%%%%%

%%%%%%%%%%%%%%%%%%%%%%%
\subsection{Decays of b-baryon}
\label{Decays of b-baryon}
%%%%%%%%%%%%%%%%%%%%%%%
First, we discuss the b-baryon decays into an octet pentaquark and a light meson. The leading-order effective Hamiltonian is given by
\begin{eqnarray}\label{Hamiltonian}
  \mathcal{H}_{\textrm{w.e.}}(b \rightarrow q c \bar{c})=\frac{G_{F}}{\sqrt{2}}\bigg(V_{c b} V_{c q}^{*}\left(C_{1} O_{1}+C_{2} O_{2}\right)\bigg) \,,
\end{eqnarray}
with
\begin{eqnarray}
  O_{1} &=& \left(\bar{c}_{\alpha} b_{\beta}\right)_{V-A}\left(\bar{q}_{\beta} c_{\alpha}\right)_{V-A} \,,\nn\\
  O_{2} &=& \left(\bar{c}_{\alpha} b_{\alpha}\right)_{V-A}\left(\bar{q}_{\beta} c_{\beta}\right)_{V-A} \,,
\end{eqnarray}
where $q$ can be $d$ or $s$. The $G_F$ and $V_{ij}$ are Fermi coupling constant and CKM matrix element respectively. $O_i$ is the low-energy effective operator and $C_i$ is the corresponding Wilson coefficient. We have neglected contributions from penguin diagrams, they are substantially suppressed compared to the tree diagrams. The operators $O_i$ transfer under the flavor SU(3) as $3$, the corresponding quark level transition $b\to c\bar c d/s$ can form an effective vertices $H$ with $(H)^{1}=0$, $(H)^{2}=V_{c d}^{*}$ and $(H)^{3}=V_{c s}^{*}$.

\begin{figure*}
\includegraphics[width=1.3\columnwidth]{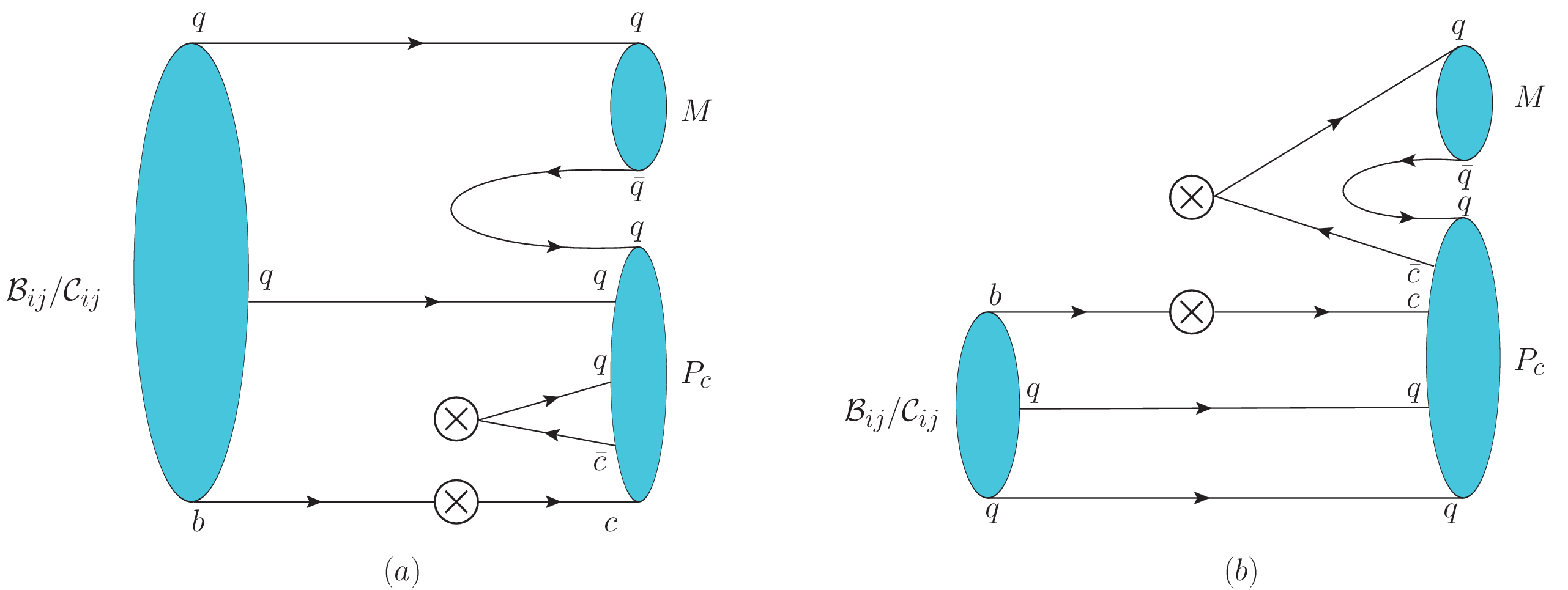}
\caption{Topological diagrams for a b-baryon decays into an octet pentaquark and a light meson. The panel (a) corresponds to the $a_{3}$, $a_{4}$, $a_{5}$ and the $b_{2}$, $b_{3}$, $b_{4}$ terms in Eq.\,(\ref{FirstbaryonH}) and (\ref{SecondbaryonH}), respectively. The panel (b) corresponds to the $a_{1}$, $a_{2}$ and the $b_{1}$ terms.}
\label{DFirstSecondbaryonH}
\end{figure*}

At the hadron level, for a b-baryon which belongs to the antitriplet decays into an octet pentaquark and a light meson, \red{the effective Hamiltonian is constructed as
\begin{eqnarray}\label{FirstbaryonH}
	{\cal H}_{\textit{eff}}&=& a_1(\mathcal{B})^{il}(H)^m \epsilon_{ijk} (\mathcal{\overline{P}})^k_l (\overline{M})^j_m \nn\\
&&+ a_2(\mathcal{B})^{im}(H)^j \epsilon_{ijk} (\mathcal{\overline{P}})^k_l (\overline{M})^l_m \nn\\
&&+ a_3(\mathcal{B})^{lm}(H)^i \epsilon_{ijk} (\mathcal{\overline{P}})^k_l (\overline{M})^j_m \,.
\end{eqnarray}
}
For a b-baryon belongs to the sextet, \red{the effective Hamiltonian is expressed as
\begin{eqnarray}\label{SecondbaryonH}
	{\cal H}_{\textit{eff}}&=& b_1(\mathcal{C})^{il}(H)^m \epsilon_{ijk} (\mathcal{\overline{P}})^k_l (\overline{M})^j_m \nn\\
&& +b_2(\mathcal{C})^{im}(H)^j \epsilon_{ijk} (\mathcal{\overline{P}})^k_l (\overline{M})^l_m \nn\\
&& +b_3(\mathcal{C})^{lm}(H)^i \epsilon_{ijk} (\mathcal{\overline{P}})^k_l (\overline{M})^j_m \,.
\end{eqnarray}
}
In the above, we suppressed the Lorentz indices and spinor forms, but only concentrate on flavor SU(3) indices. Here the $a_i$ and $b_i$ are SU(3) irreducible nonperturbative amplitudes. Topological diagrams for these decay modes are given in Fig.\,\ref{DFirstSecondbaryonH}. The individual decay amplitude can be obtained by expanding Eqs.\,(\ref{FirstbaryonH}) and (\ref{SecondbaryonH}), they are collected in Tables\,\ref{tab:P8_1} and \ref{tab:P8_2}. From these results, one can read off much information. We present some of the interesting properties in the following.

%%%%%%%%%%%%%%%%%%%%%%
%%%%%%%%%%%%%%%%%%%%%%

\begin{table*}
	\setlength{\tabcolsep}{18pt}
	\caption{Amplitudes for b-baryon (antitriplet) decays into a pentaquark and a light meson.}\label{tab:P8_1}\renewcommand{\arraystretch}{1.6}
  \begin{tabular}{l c l c}\hline
		channel & amplitude & channel & amplitude\\\hline
		$\Lambda_b^0\to   P_{\Sigma^-}  \pi^+  $ & $ \left(a_3-a_2\right) V_{\text{cs}}^*$&
		$\Lambda_b^0\to   P_{\Sigma^-}  K^+  $ & $ -a_3 V_{\text{cd}}^*$\\
		$\Lambda_b^0\to   P_{\Sigma^0}  \pi^0  $ & $ \left(a_3-a_2\right) V_{\text{cs}}^*$&
		$\Lambda_b^0\to   P_{\Sigma^0}  K^0  $ & $ \frac{a_3}{\sqrt{2}}V_{\text{cd}}^*$\\
		$\Lambda_b^0\to   P_{\Sigma^+}  \pi^-  $ & $ \left(a_3-a_2\right) V_{\text{cs}}^*$&
		$\Lambda_b^0\to   P_{p}  \pi^-  $ & $ \left(a_1+a_2-a_3\right) V_{\text{cd}}^*$\\
		$\Lambda_b^0\to   P_{p}  K^-  $ & $ a_1 V_{\text{cs}}^*$&
		$\Lambda_b^0\to   P_{n}  \pi^0  $ & $ -\frac{\left(a_1+a_2-a_3\right) }{\sqrt{2}}V_{\text{cd}}^*$\\
		$\Lambda_b^0\to   P_{n}  \overline K^0  $ & $ a_1 V_{\text{cs}}^*$&
		$\Lambda_b^0\to   P_{\Lambda}  K^0  $ & $ -\frac{\left(2 a_1+2 a_2-a_3\right) }{\sqrt{6}}V_{\text{cd}}^*$\\
		$\Xi_b^0\to   P_{\Lambda}  \overline K^0  $ & $ -\frac{\left(a_1+a_2+a_3\right) }{\sqrt{6}}V_{\text{cs}}^*$&
		$\Xi_b^0\to   P_{\Sigma^0}  \pi^0  $ & $ \frac{\left(-a_1+a_2\right)}{2}  V_{\text{cd}}^*$\\
		$\Xi_b^0\to   P_{\Sigma^0}  \overline K^0  $ & $ \frac{\left(a_1+a_2-a_3\right) }{\sqrt{2}}V_{\text{cs}}^*$&
		$\Xi_b^0\to   P_{\Sigma^+}  \pi^-  $ & $ -a_1 V_{\text{cd}}^*$\\
		$\Xi_b^0\to   P_{\Sigma^+}  K^-  $ & $ -\left(a_1+a_2-a_3\right) V_{\text{cs}}^*$&
		$\Xi_b^0\to   P_{p}  K^-  $ & $ \left(a_2-a_3\right) V_{\text{cd}}^*$\\
		$\Xi_b^-\to   P_{\Lambda}  K^-  $ & $ \frac{\left(a_1+a_2+a_3\right) }{\sqrt{6}}V_{\text{cs}}^*$&
		$\Xi_b^0\to   P_{\Lambda}  \pi^0  $ & $ \frac{\left(a_1+a_2-2 a_3\right) }{2 \sqrt{3}}V_{\text{cd}}^*$\\
		$\Xi_b^-\to   P_{\Sigma^-}  \overline K^0  $ & $ \left(a_1+a_2-a_3\right) V_{\text{cs}}^*$&
		$\Xi_b^0\to   P_{n}  \overline K^0  $ & $ a_2 V_{\text{cd}}^*$\\
		$\Xi_b^-\to   P_{\Sigma^0}  K^-  $ & $ \frac{\left(a_1+a_2-a_3\right) }{\sqrt{2}}V_{\text{cs}}^*$&
		$\Xi_b^-\to   P_{\Lambda}  \pi^-  $ & $ \frac{\left(a_1+a_2-2 a_3\right) }{\sqrt{6}}V_{\text{cd}}^*$\\&&
		$\Xi_b^-\to   P_{\Sigma^-}  \pi^0  $ & $ -\frac{\left(a_1+a_2\right)}{\sqrt{2}} V_{\text{cd}}^*$\\&&
		$\Xi_b^0\to P_{\Sigma^-} \pi^+ $ & $  a_2 V_{\text{cd}}^*$\\&&
		$\Xi_b^-\to   P_{n}  K^-  $ & $ -a_3 V_{\text{cd}}^*$\\&&
		$\Xi_b^-\to   P_{\Sigma^0}  \pi^-  $ & $ \frac{\left(a_1+a_2\right) }{\sqrt{2}}V_{\text{cd}}^*$\\\hline
  \end{tabular}
\end{table*}

\begin{table*}	
  \setlength{\tabcolsep}{18pt}
  \caption{Amplitudes for b-baryon (sextet) decays into a pentaquark and a light meson.}\label{tab:P8_2}\renewcommand{\arraystretch}{1.6}\begin{tabular}{l c l c}\hline
			channel & amplitude & channel & amplitude\\\hline
			$\Sigma_{b}^{+}\to   P_{\Lambda}  \pi^+  $ & $ -\frac{\left(b_2+b_3\right)}{\sqrt{6}} V_{\text{cs}}^*$&
			$\Sigma_{b}^{+}\to   P_{\Lambda}  K^+  $ & $ \frac{\left(-2 b_2+b_3\right)}{\sqrt{6}} V_{\text{cd}}^*$\\
			$\Sigma_{b}^{+}\to   P_{\Sigma^0}  \pi^+  $ & $ \frac{\left(b_2-b_3\right) }{\sqrt{2}}V_{\text{cs}}^*$&
			$\Sigma_{b}^{+}\to   P_{\Sigma^0}  K^+  $ & $ \frac{b_3}{\sqrt{2}}V_{\text{cd}}^*$\\
			$\Sigma_{b}^{+}\to   P_{\Sigma^+}  \pi^0  $ & $ \frac{\left(b_3-b_2\right) }{\sqrt{2}}V_{\text{cs}}^*$&
			$\Sigma_{b}^{+}\to   P_{\Sigma^+}  K^0  $ & $ -b_1 V_{\text{cd}}^*$\\
			$\Sigma_{b}^{+}\to   P_{p}  \overline K^0  $ & $ b_1 V_{\text{cs}}^*$&
			$\Sigma_{b}^{+}\to   P_{n}  \pi^+  $ & $ b_2 V_{\text{cd}}^*$\\
			$\Sigma_{b}^{0}\to   P_{\Lambda}  \pi^0  $ & $ \frac{\left(b_2+b_3\right) }{\sqrt{6}}V_{\text{cs}}^*$&
			$\Sigma_{b}^{0}\to   P_{\Lambda}  K^0  $ & $ -\frac{\left(2 b_2-b_3\right) }{2 \sqrt{3}}V_{\text{cd}}^*$\\
			$\Sigma_{b}^{0}\to   P_{\Sigma^-}  \pi^+  $ & $ \frac{\left(b_2-b_3\right) }{\sqrt{2}}V_{\text{cs}}^*$&
			$\Sigma_{b}^{0}\to   P_{\Sigma^-}  K^+  $ & $ \frac{b_3}{\sqrt{2}}V_{\text{cd}}^*$\\
			$\Sigma_{b}^{0}\to   P_{\Sigma^+}  \pi^-  $ & $ \frac{\left(b_3-b_2\right) }{\sqrt{2}}V_{\text{cs}}^*$&
			$\Sigma_{b}^{+}\to   P_{p}  \pi^0  $ & $ -\frac{\left(b_1-b_2+b_3\right) }{\sqrt{2}}V_{\text{cd}}^*$\\
			$\Sigma_{b}^{0}\to   P_{p}  K^-  $ & $ -\frac{b_1}{\sqrt{2}}V_{\text{cs}}^*$&
			$\Sigma_{b}^{0}\to   P_{\Sigma^0}  K^0  $ & $ \frac{\left(2 b_1+b_3\right)}{2}  V_{\text{cd}}^*$\\
			$\Sigma_{b}^{0}\to   P_{n}  \overline K^0  $ & $ \frac{b_1}{\sqrt{2}}V_{\text{cs}}^*$&
			$\Sigma_{b}^{0}\to   P_{p}  \pi^-  $ & $ -\frac{\left(b_1-b_2+b_3\right) }{\sqrt{2}}V_{\text{cd}}^*$\\
			$\Sigma_{b}^{-}\to   P_{\Lambda}  \pi^-  $ & $ \frac{\left(b_2+b_3\right) }{\sqrt{6}}V_{\text{cs}}^*$&
			$\Sigma_{b}^{0}\to   P_{n}  \pi^0  $ & $ -\frac{\left(b_1+b_2+b_3\right)}{2}  V_{\text{cd}}^*$\\
			$\Sigma_{b}^{-}\to   P_{\Sigma^-}  \pi^0  $ & $ \frac{\left(b_3-b_2\right) }{\sqrt{2}}V_{\text{cs}}^*$&
			$\Sigma_{b}^{-}\to   P_{\Sigma^-}  K^0  $ & $ \left(b_1+b_3\right) V_{\text{cd}}^*$\\
			$\Sigma_{b}^{-}\to   P_{\Sigma^0}  \pi^-  $ & $ \frac{\left(b_2-b_3\right) }{\sqrt{2}}V_{\text{cs}}^*$&
			$\Sigma_{b}^{-}\to   P_{n}  \pi^-  $ & $ -\left(b_1+b_3\right) V_{\text{cd}}^*$\\
			$\Sigma_{b}^{-}\to   P_{n}  K^-  $ & $ -b_1 V_{\text{cs}}^*$&
			$\Xi_{b}^{\prime0}\to   P_{\Lambda}  \pi^0  $ & $ \frac{\left(3 b_1-b_2+2 b_3\right) }{2 \sqrt{6}}V_{\text{cd}}^*$\\
			$\Xi_{b}^{\prime0}\to   P_{\Lambda}  \overline K^0  $ & $ -\frac{\left(3 b_1+b_2+b_3\right) }{2 \sqrt{3}}V_{\text{cs}}^*$&
			$\Xi_{b}^{\prime0}\to   P_{\Sigma^-}  \pi^+  $ & $ -\frac{b_2}{\sqrt{2}}V_{\text{cd}}^*$\\
			$\Xi_{b}^{\prime0}\to   P_{\Sigma^0}  \overline K^0  $ & $ -\frac{\left(b_1-b_2+b_3\right)}{2}  V_{\text{cs}}^*$&
			$\Xi_{b}^{\prime0}\to   P_{\Sigma^0}  \pi^0  $ & $ \frac{\left(b_1-b_2\right) }{2 \sqrt{2}}V_{\text{cd}}^*$\\
			$\Xi_{b}^{\prime0}\to   P_{\Sigma^+}  K^-  $ & $ \frac{\left(b_1-b_2+b_3\right) }{\sqrt{2}}V_{\text{cs}}^*$&
			$\Xi_{b}^{\prime0}\to   P_{\Sigma^+}  \pi^-  $ & $ \frac{b_1}{\sqrt{2}}V_{\text{cd}}^*$\\
			$\Xi_{b}^{\prime-}\to   P_{\Lambda}  K^-  $ & $ \frac{\left(3 b_1+b_2+b_3\right) }{2 \sqrt{3}}V_{\text{cs}}^*$&
			$\Xi_{b}^{\prime0}\to   P_{p}  K^-  $ & $ \frac{\left(b_2-b_3\right) }{\sqrt{2}}V_{\text{cd}}^*$\\
			$\Xi_{b}^{\prime-}\to   P_{\Sigma^-}  \overline K^0  $ & $ -\frac{\left(b_1-b_2+b_3\right) }{\sqrt{2}}V_{\text{cs}}^*$&
			$\Xi_{b}^{\prime0}\to   P_{n}  \overline K^0  $ & $ \frac{b_2}{\sqrt{2}}V_{\text{cd}}^*$\\
			$\Xi_{b}^{\prime-}\to   P_{\Sigma^0}  K^-  $ & $ -\frac{\left(b_1-b_2+b_3\right)}{2}  V_{\text{cs}}^*$&
			$\Xi_{b}^{\prime-}\to   P_{\Lambda}  \pi^-  $ & $ \frac{\left(3 b_1-b_2+2 b_3\right) }{2 \sqrt{3}}V_{\text{cd}}^*$\\&&
			$\Xi_{b}^{\prime-}\to   P_{\Sigma^-}  \pi^0  $ & $ \frac{\left(b_1+b_2\right)}{2}  V_{\text{cd}}^*$\\&&
			$\Xi_{b}^{\prime-}\to   P_{\Sigma^0}  \pi^-  $ & $ -\frac{\left(b_1+b_2\right)}{2}  V_{\text{cd}}^*$\\&&
			$\Xi_{b}^{\prime-}\to   P_{n}  K^-  $ & $ -\frac{b_3}{\sqrt{2}}V_{\text{cd}}^*$\\&&
			$\Omega_{b}^{-}\to   P_{\Lambda}  K^-  $ & $ \frac{\left(-b_2+2 b_3\right) }{\sqrt{6}}V_{\text{cd}}^*$\\&&
			$\Omega_{b}^{-}\to   P_{\Sigma^-}  \overline K^0  $ & $ -b_2 V_{\text{cd}}^*$\\&&
			$\Omega_{b}^{-}\to   P_{\Sigma^0}  K^-  $ & $ -\frac{b_2}{\sqrt{2}}V_{\text{cd}}^*$\\\hline
  \end{tabular}
\end{table*}

\begin{enumerate}
  \item Table\,\ref{tab:P8_1} and \ref{tab:P8_2} are arranged according to the dependence on CKM matrix elements, $c\to s$ transition is proportional to $|V_{cs}^*|\sim 1$; while $c\to d$ transition is Cabibbo suppressed $|V_{cd}^*|\sim 0.2$.

  \item A number of relations for different decay widths can be readily read off from Table\,\ref{tab:P8_1}:
  \begin{eqnarray}
    \bm{ \Gamma(\Lambda_b^0\to P_{p} K^- ) = \Gamma(\Lambda_b^0\to P_{n} \overline K^0 ) \,,} \nn\\
    \bm{ \Gamma(\Lambda_b^0\to P_{p} \pi^- ) = 2\Gamma(\Lambda_b^0\to P_{n} \pi^0 ) \,,}\nn\\
    \bm{ \Gamma(\Lambda_b^0\to P_{\Sigma^-} K^+ ) = 2\Gamma(\Lambda_b^0\to P_{\Sigma^0} K^0 ) \,,} \nn\\
    \bm{ \Gamma(\Lambda_b^0\to P_{\Sigma^-} \pi^+ ) = \Gamma(\Lambda_b^0\to P_{\Sigma^+} \pi^- ) } \nn\\
    \bm{ =\Gamma(\Lambda_b^0\to P_{\Sigma^0} \pi^0 ) \,,}\nn\\
    \bm{ \Gamma(\Xi_b^0\to P_{\Lambda} \overline K^0 ) = \Gamma(\Xi_b^-\to P_{\Lambda} K^- ) \,,} \nn\\
    \bm{ \Gamma(\Xi_b^-\to P_{\Sigma^-} \pi^0 ) = \Gamma(\Xi_b^-\to P_{\Sigma^0} \pi^- ) \,,} \nn\\
    \bm{ \Gamma(\Xi_b^-\to P_{\Lambda} \pi^- ) = 2\Gamma(\Xi_b^0\to P_{\Lambda} \pi^0 ) \,,} \nn\\
    \bm{ \Gamma(\Xi_b^-\to P_{\Sigma^-} \overline K^0 ) = \Gamma(\Xi_b^0\to P_{\Sigma^+} K^- ) } \nn\\
    \bm{ =2\Gamma(\Xi_b^0\to P_{\Sigma^0} \overline K^0 ) \,,} \nn\\
    \bm{ =2\Gamma(\Xi_b^-\to P_{\Sigma^0} K^- ) \,,} \nn\\
    \Gamma(\Lambda_b^0\to P_{\Sigma^-} K^+ ) =  \Gamma(\Xi_b^-\to P_{n} K^- ) \,,\nn\\
	\Gamma(\Xi_b^0\to P_{\Sigma^-} \pi^+ ) = \Gamma(\Xi_b^0\to P_{n} \overline K^0 ) \,.
  \end{eqnarray}
  And the relations deduced from Table\,\ref{tab:P8_2}:
  \begin{eqnarray}
    \bm{ \Gamma(\Omega_{b}^{-}\to P_{\Sigma^-} \overline K^0 ) = 2\Gamma(\Omega_{b}^{-}\to P_{\Sigma^0} K^- ) \,,}\nn\\
    \bm{ \Gamma(\Xi_{b}^{\prime0}\to P_{\Lambda} \overline K^0 ) = \Gamma(\Xi_{b}^{\prime-}\to P_{\Lambda} K^- )  \,,}\nn\\
    \bm{ \Gamma(\Xi_{b}^{\prime0}\to P_{\Sigma^+} K^- ) = \Gamma(\Xi_{b}^{\prime-}\to P_{\Sigma^-} \overline K^0 )  }\nn\\
    \bm{ = 2\Gamma(\Xi_{b}^{\prime0}\to P_{\Sigma^0} \overline K^0 ) }\nn\\
    \bm{ = 2\Gamma(\Xi_{b}^{\prime-}\to P_{\Sigma^0} K^- ) \,,}\nn\\
    \bm{ \Gamma(\Xi_{b}^{\prime-}\to P_{\Lambda} \pi^- ) = 2\Gamma(\Xi_{b}^{\prime0}\to P_{\Lambda} \pi^0 ) \,,}\nn\\
    \bm{ \Gamma(\Xi_{b}^{\prime-}\to P_{\Sigma^-} \pi^0 ) = \Gamma(\Xi_{b}^{\prime-}\to P_{\Sigma^0} \pi^- ) \,,}\nn\\
    \bm{ \Gamma(\Sigma_{b}^{+}\to P_{\Lambda} \pi^+ ) = \Gamma(\Sigma_{b}^{-}\to P_{\Lambda} \pi^- )  \,,}\nn\\
    \bm{ \Gamma(\Sigma_{b}^{+}\to P_{\Sigma^0} \pi^+ ) = \Gamma(\Sigma_{b}^{-}\to P_{\Sigma^0} \pi^- )  \,,}\nn\\
    \bm{ \Gamma(\Sigma_{b}^{+}\to P_{p} \overline K^0 ) = \Gamma(\Sigma_{b}^{-}\to P_{n} K^- )  \,,}\nn\\
    \bm{ \Gamma(\Sigma_{b}^{+}\to P_{\Lambda} \pi^+ ) = \Gamma(\Sigma_{b}^{0}\to P_{\Lambda} \pi^0 ) \,,}\nn\\
    \bm{ \Gamma(\Sigma_{b}^{+}\to P_{\Lambda} K^+ ) = 2\Gamma(\Sigma_{b}^{0}\to P_{\Lambda} K^0 ) \,,}\nn\\
    \bm{ \Gamma(\Sigma_{b}^{+}\to P_{\Sigma^0} K^+ ) = \Gamma(\Sigma_{b}^{0}\to P_{\Sigma^-} K^+ ) \,,}\nn\\
    \bm{ \Gamma(\Sigma_{b}^{+}\to P_{p} \pi^0 ) = \Gamma(\Sigma_{b}^{0}\to P_{p} \pi^- ) \,,}\nn\\
    \bm{ \Gamma(\Sigma_{b}^{+}\to P_{p} \overline K^0 ) = 2\Gamma(\Sigma_{b}^{0}\to P_{n} \overline K^0 ) }\nn\\
    \bm{ = 2\Gamma(\Sigma_{b}^{0}\to P_{p} K^- ) \,,}\nn\\
    \bm{ \Gamma(\Sigma_{b}^{+}\to P_{\Sigma^0} \pi^+ ) = \Gamma(\Sigma_{b}^{+}\to P_{\Sigma^+} \pi^0 ) }\nn\\
    \bm{ = \Gamma(\Sigma_{b}^{0}\to P_{\Sigma^+} \pi^- ) }\nonumber\\
    \bm{ = \Gamma(\Sigma_{b}^{0}\to P_{\Sigma^-} \pi^+ ) }\nonumber\\
    \bm{ = \Gamma(\Sigma_{b}^{-}\to P_{\Sigma^-} \pi^0 ) \,,}\nn\\
    \bm{ \Gamma(\Sigma_{b}^{0}\to P_{\Lambda} \pi^0 ) = \Gamma(\Sigma_{b}^{-}\to P_{\Lambda} \pi^- ) \,,}\nn\\\nn
  \end{eqnarray}
  \begin{eqnarray}
    \Gamma(\Omega_{b}^{-}\to P_{\Sigma^-} \overline K^0 ) &=& \Gamma(\Sigma_{b}^{+}\to P_{n} \pi^+ ) \nonumber\\
    &=& 2\Gamma(\Xi_{b}^{\prime0}\to P_{n} \overline K^0 )\nonumber\\
    &=& 2\Gamma(\Xi_{b}^{\prime0}\to P_{\Sigma^-} \pi^+ )  \,,\nonumber\\
    \Gamma(\Sigma_{b}^{+}\to P_{\Sigma^0} K^+ ) &=& \Gamma(\Xi_{b}^{\prime-}\to P_{n} K^- ) \nonumber\\
    \Gamma(\Sigma_{b}^{+}\to P_{\Sigma^+} K^0 ) &=& 2\Gamma(\Xi_{b}^{\prime0}\to P_{\Sigma^+} \pi^- ) \,,\nn\\
	\Gamma(\Sigma_{b}^{-}\to P_{\Sigma^-} K^0 ) &=& \Gamma(\Sigma_{b}^{-}\to P_{n} \pi^- ) \,.
  \end{eqnarray}	
    The relations marked in bold are upheld by I-spin symmetry, which are more reliable than the U- and V-spin relations. The results show above just give the relative relations, absolute decay rates require a reliable computation of the irreducible nonperturbative amplitudes. This, however, is a daunting task, way beyond the theoretical methods available presently. Besides, the decay modes of $\Omega_b^-$ might be experimentally more important since the decays of $\Sigma_b$ and $\Xi_{b}$ are dominated by strong interactions.

  \item If we take $P_c(4312)$ as $P_p$ and $P_{cs}(4459)$ as $P_{\Lambda}$ in Eq.\,(\ref{multiPc}) according to the different light valence quark components. Then some of these relations shown above would provide hints for the exploration of new decay modes. Combined with the strong decays of pentaqurk, we will list some cascade decay modes, which are likely to be utilized to reconstruct the pentaquark, in the next section.
\end{enumerate}

However, it is necessary to point out the above relationships between decay widths are only a ball-park estimate, they are obtained in the flavor SU(3) symmetry limit in which the mass differences between final state hadrons have been ignored. In addition, the hadronization processes would also influence the relations derived in this work. Although the SU(3) breaking effects might be sizable, our qualitative results should be relatively robust, unless the flavor symmetry is broken much strongly in bottom quark decay than empirically anticipated. With more data from LCHb and other experiments, a rigorous analysis would be necessary in future \cite{Xing:2023bzh,Geng:2019bfz}.

%%%%%%%%%%%%%%%%%%%%%%%
\subsection{Decays of B-meson}
\label{Decays of B-meson}
%%%%%%%%%%%%%%%%%%%%%%%
At the hadron level, for a B-meson which belongs to an SU(3) antitriplet decays into an octet pentaquark and a light antibaryon, the corresponding effective Hamiltonian is constructed as
\begin{eqnarray}\label{mesonH}
	{\cal H}_{\textit{eff}}&=& c_1 (B)_n (H)^n \epsilon_{ijk} (\mathcal{\overline{P}})^k_l \epsilon^{ilm} (T_8)^j_m \nn\\
&&+ c_2 (B)_n (H)^l \epsilon_{ijk} (\mathcal{\overline{P}})^k_l \epsilon^{inm} (T_8)^j_m \nn\\
&&+ c_3 (B)_n (H)^j \epsilon_{ijk} (\mathcal{\overline{P}})^k_l \epsilon^{inm} (T_8)^l_m \,.
\end{eqnarray}
The topological diagrams for these decays are given in Fig.\,\ref{DmesonH}. The decay amplitudes for different channels can be deduced from the Hamiltonian in Eq.\,(\ref{mesonH}), they are displayed in Table\,\ref{tab:P8_3}. From these amplitudes, we can find the relations for decay widths in the SU(3) symmetry limit:

%%%%%%%%%%%%%%%%%%%%%%
\begin{figure*}
\includegraphics[width=1.3\columnwidth]{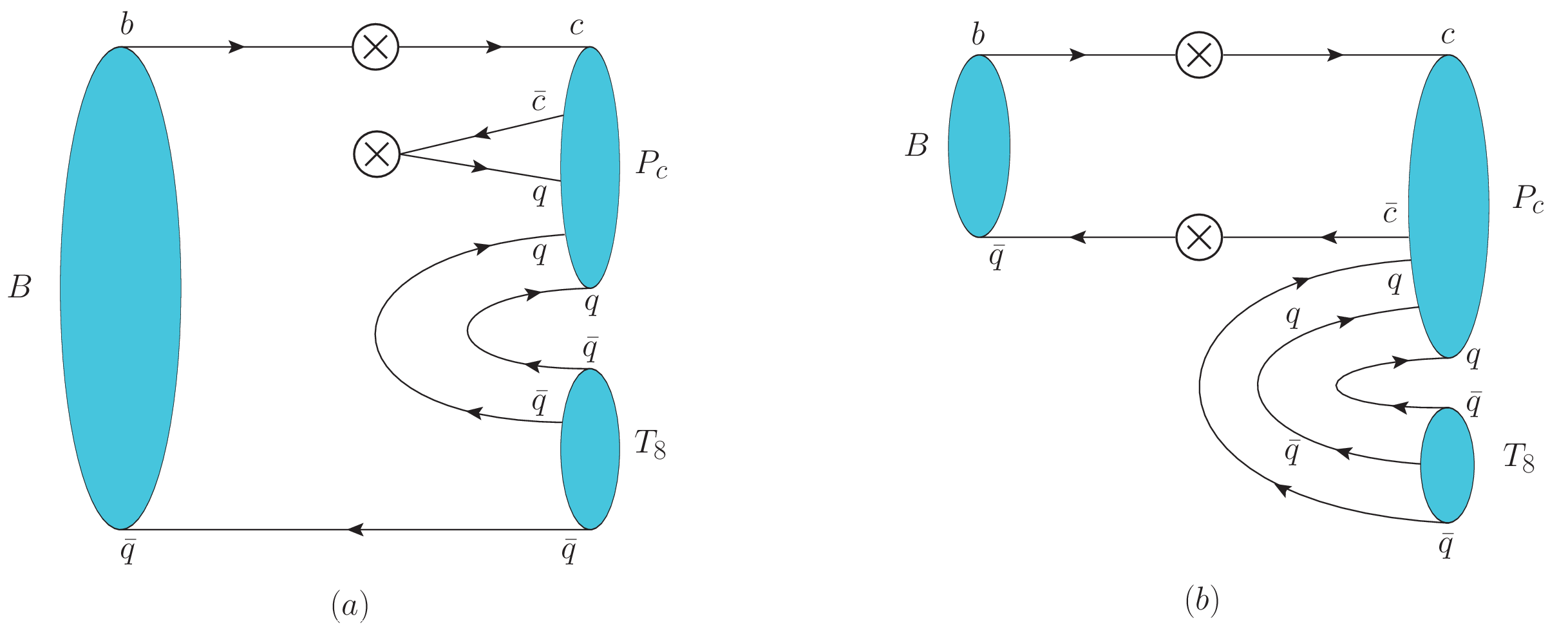}
\caption{Topological diagrams for a B-meson decays into an octet pentaquark and a light antibaryon. The panel (a) refers to the $c_{2}$ and $c_{3}$ terms and the panel (b) to $c_{1}$ term in Eq.\,(\ref{mesonH}).}
\label{DmesonH}
\end{figure*}
%%%%%%%%%%%%%%%%%%%%%%

\begin{table*}
  \setlength{\tabcolsep}{8pt}
  \caption{Amplitudes for B-meson decays into a pentaquark and a light baryon.}\label{tab:P8_3}\renewcommand{\arraystretch}{1.6}\begin{tabular}{l c l c}\hline channel & amplitude & channel & amplitude\\\hline
			$B^-\to   P_{\Lambda}  \overline p $ & $ -\frac{\left(2 c_2+c_3\right) }{\sqrt{6}}V_{\text{cs}}^*$& $
			B^-\to   P_{\Sigma^-}  \overline \Lambda^0 $ & $ \frac{\left(c_2-c_3\right) }{\sqrt{6}}V_{\text{cd}}^*$\\
			$B^-\to   P_{\Sigma^-}  \overline n $ & $ -c_3 V_{\text{cs}}^*$& $
			B^-\to   P_{\Lambda}  \overline \Sigma^- $ & $ \frac{\left(c_2-c_3\right) }{\sqrt{6}}V_{\text{cd}}^*$\\
			$B^-\to   P_{\Sigma^0}  \overline p $ & $ -\frac{c_3}{\sqrt{2}}V_{\text{cs}}^*$& $
			B^-\to   P_{\Sigma^-}  \overline \Sigma^0 $ & $ \frac{\left(c_2+c_3\right) }{\sqrt{2}}V_{\text{cd}}^*$\\
			$\overline B^0\to   P_{\Lambda}  \overline n $ & $ -\frac{\left(2 c_2+c_3\right) }{\sqrt{6}}V_{\text{cs}}^*$& $
			B^-\to   P_{\Sigma^0}  \overline \Sigma^- $ & $ -\frac{\left(c_2+c_3\right) }{\sqrt{2}}V_{\text{cd}}^*$\\
			$\overline B^0\to   P_{\Sigma^0}  \overline n $ & $ \frac{c_3}{\sqrt{2}}V_{\text{cs}}^*$& $
			B^-\to   P_{n}  \overline p $ & $ c_2 V_{\text{cd}}^*$\\
			$\overline B^0\to   P_{\Sigma^+}  \overline p $ & $ -c_3 V_{\text{cs}}^*$& $
			\overline B^0\to   P_{\Lambda}  \overline \Lambda^0 $ & $ \frac{\left(6 c_1+c_2+5 c_3\right)}{6}  V_{\text{cd}}^*$\\
			$\overline B^0_s\to   P_{\Lambda}  \overline \Lambda^0 $ & $ \frac{\left(3 c_1+2 c_2+c_3\right)}{3}  V_{\text{cs}}^*$& $
			\overline B^0\to   P_{\Lambda}  \overline \Sigma^0 $ & $ -\frac{\left(c_2-c_3\right) }{2 \sqrt{3}}V_{\text{cd}}^*$\\
			$\overline B^0_s\to   P_{\Sigma^-}  \overline \Sigma^+ $ & $ \left(c_1+c_3\right) V_{\text{cs}}^*$& $
			\overline B^0\to   P_{\Sigma^-}  \overline \Sigma^+ $ & $ \left(c_1+c_2+c_3\right) V_{\text{cd}}^*$\\
			$\overline B^0_s\to   P_{\Sigma^0}  \overline \Sigma^0 $ & $ \left(c_1+c_3\right) V_{\text{cs}}^*$& $
			\overline B^0\to   P_{\Sigma^0}  \overline \Lambda^0 $ & $ -\frac{\left(c_2-c_3\right)}{2 \sqrt{3}} V_{\text{cd}}^*$\\
			$\overline B^0_s\to   P_{\Sigma^+}  \overline \Sigma^- $ & $ \left(c_1+c_3\right) V_{\text{cs}}^*$& $
			\overline B^0\to   P_{\Sigma^0}  \overline \Sigma^0 $ & $ \frac{\left(2 c_1+c_2+c_3\right)}{2}  V_{\text{cd}}^*$\\
			$\overline B^0_s\to   P_{p}  \overline p $ & $ c_1 V_{\text{cs}}^*$& $
			\overline B^0\to   P_{\Sigma^+}  \overline \Sigma^- $ & $ c_1 V_{\text{cd}}^*$\\
			$\overline B^0_s\to   P_{n}  \overline n $ & $ c_1 V_{\text{cs}}^*$& $
			\overline B^0\to   P_{p}  \overline p $ & $ \left(c_1+c_2\right) V_{\text{cd}}^*$\\&&
			$\overline B^0\to   P_{n}  \overline n $ & $ \left(c_1+c_2+c_3\right) V_{\text{cd}}^*$\\&&
			$\overline B^0_s\to   P_{\Lambda}  \overline \Xi^0 $ & $ \frac{\left(c_2+2 c_3\right) }{\sqrt{6}}V_{\text{cd}}^*$\\&&
			$\overline B^0_s\to   P_{\Sigma^-}  \overline \Xi^+ $ & $ c_2 V_{\text{cd}}^*$\\&&
			$\overline B^0_s\to   P_{\Sigma^0}  \overline \Xi^0 $ & $ -\frac{c_2}{\sqrt{2}}V_{\text{cd}}^*$\\&&
			$\overline B^0_s\to   P_{p}  \overline \Sigma^- $ & $ -c_3 V_{\text{cd}}^*$\\&&
			$\overline B^0_s\to   P_{n}  \overline \Lambda^0 $ & $ -\frac{\left(2 c_2+c_3\right) }{\sqrt{6}}V_{\text{cd}}^*$\\&&
			$\overline B^0_s\to   P_{n}  \overline \Sigma^0 $ & $ \frac{c_3}{\sqrt{2}} V_{\text{cd}}^*$\\\hline
  \end{tabular}
\end{table*}

\begin{eqnarray}\label{relationBmeson}
  \Gamma(B^-\to P_{\Lambda} \overline \Sigma^-) &=& \Gamma(B^-\to P_{\Sigma^-} \overline \Lambda^0) \nonumber\\
  &=& 2\Gamma(\overline B^0\to P_{\Lambda} \overline \Sigma^0) \nonumber\\
  &=& 2\Gamma(\overline B^0\to P_{\Sigma^0} \overline \Lambda^0) \,,\nonumber\\
  \Gamma(B^-\to   P_{\Sigma^-}  \overline \Sigma^0) &=& \Gamma	(B^-\to   P_{\Sigma^0}  \overline \Sigma^- ) \,,\nonumber\\
  \Gamma(B^-\to P_{\Sigma^-} \overline n) &=& \Gamma(\overline B^0\to P_{\Sigma^+} \overline p) \nonumber\\
  &=& 2\Gamma(\overline B^0\to P_{\Sigma^0} \overline n) \nonumber\\
  &=& 2\Gamma(B^-\to P_{\Sigma^0} \overline p) \,,\nonumber
\end{eqnarray}
\begin{eqnarray}
  \Gamma(B^-\to P_{\Lambda} \overline p) &=& \Gamma(\overline B^0\to P_{\Lambda} \overline n) \,,\nonumber\\
  \Gamma(\overline B^0_s\to P_{p} \overline \Sigma^-) &=& 2\Gamma(\overline B^0_s\to P_{n} \overline \Sigma^0) \,,\nonumber\\
  \Gamma(\overline B^0\to P_{\Sigma^-} \overline \Sigma^+) &=& \Gamma(\overline B^0\to P_{n} \overline n) \,,\nonumber\\
  \Gamma(\overline B^0_s\to P_{\Sigma^-} \overline \Sigma^+) &=& \Gamma(\overline B^0_s\to P_{\Sigma^0} \overline \Sigma^0) \nonumber\\
  &=& \Gamma(\overline B^0_s\to P_{\Sigma^+} \overline \Sigma^-) \,,\nonumber\\
  \Gamma(\overline B^0_s\to P_{\Sigma^-} \overline \Xi^+) &=& \Gamma(B^-\to   P_{n}  \overline p) \nonumber\\
  &=& 2\Gamma(\overline B^0_s\to P_{\Sigma^0} \overline \Xi^0) \,,\nonumber\\
  \Gamma(\overline B^0_s\to P_{p} \overline p) &=& \Gamma(\overline B^0_s\to P_{n} \overline n) \,.
\end{eqnarray}

Amplitude analyses of $B_s^0 \to J/\psi \, p \, \bar p$ and $B^- \to J/\psi \Lambda \, \overline{p}$ were performed by LHCb Collaboration recently, and evidences for charmonium pentaquarks were found. Unlike the baryonic decay, the mesonic decay offers a cleaner environment to search for new pentaquark. The relations in Eq.\,(\ref{relationBmeson}) can be utilized to find new decay channels; for instance, the Cabibbo-allowed processes $\overline B^0\to P_{\Lambda} \overline n$, $\overline B^0\to   P_{\Sigma^+}  \overline p $ and $\overline B^0_s\to   P_{n}  \overline n $ have the potential to be experimentally discovered in future.

%%%%%%%%%%%%%%%%%%%%%%%
\section{Strong decay of pentaquark}
\label{Strongdecay}
%%%%%%%%%%%%%%%%%%%%%%%
Particular decay processes of $P_c$ states in the detectors can be adapted as signatures to reconstruct these exotic states. Currently, the experimental searches for pentaquark mainly focus on the strong decays of $P_c$, like the discoveries of $P_c(4312) \to J/\psi \, p$ \cite{LHCb:2019kea} and $P_{cs}(4459) \to J/\psi \Lambda$ \cite{LHCb:2020jpq}. The effective Hamiltonian for an octet pentaquark decays into $J/\psi$ plus a light baryon is given as
\begin{eqnarray}\label{Strongdecay1}
	{\cal H}_{\textit{eff}}&=& d_1 \epsilon^{ijk} (\mathcal{P})^l_k \epsilon_{ilm} (\overline{T}_8)^m_j J/\psi \,.
\end{eqnarray}
These processes belong to strong decay, thus there is no effective vertices, this is a unique property comparing with weak decays of b-baryon and B-meson. The decay amplitudes deduced from Eq.\,(\ref{Strongdecay1}) are given in Table \ref{tab:P8_4}, which indicates all the decay widths are same:
\begin{eqnarray}
	\Gamma(P_{\Lambda}\to \Lambda^0 J/\psi)&=& \Gamma(P_{\Sigma^-}\to \Sigma^- J/\psi) \nonumber\\ &=& \Gamma(P_{\Sigma^0}\to \Sigma^0 J/\psi) \nonumber\\&=& \Gamma(P_{\Sigma^+}\to \Sigma^+ J/\psi) \nonumber\\&=& \Gamma(P_{p}\to p J/\psi) \nonumber\\&=&
\Gamma(P_{n}\to n J/\psi) \,.
\end{eqnarray}

Other possible processes include an octet pentaquark decays into anticharmed meson plus singly charmed baryon in antitriplet or sextet.
\begin{eqnarray}\label{Strongdecay2}
	{\cal H}_{\textit{eff}}&=& e_1 \epsilon^{ijk} (\mathcal{P})^l_k (\overline{T}_{\bf{c\bar 3}})_{il} D_j \nn\\
&&+ e_2 \epsilon^{ijk} (\mathcal{P})^l_k (\overline{T}_{\bf{c 6}})_{il} D_j \,.
\end{eqnarray}
The corresponding decay amplitudes are given in Table \ref{tab:P8_4}, which gives the relations between various decay widths£º
\begin{eqnarray}
 2\Gamma(P_{p}\to \Lambda_c^+ \overline D^0)  &=&2\Gamma(P_{n}\to \Lambda_c^+ D^-)  \nonumber\\
  &=& 2\Gamma(P_{\Sigma^-}\to \Xi_c^0 D^-) \nonumber\\
  &=& 2\Gamma(P_{\Sigma^+}\to \Xi_c^+ \overline D^0) \nonumber\\
  &=& 3\Gamma(P_{\Lambda}\to \Lambda_c^+  D^-_s) \nonumber\\
  &=& 4\Gamma(P_{\Sigma^0}\to \Xi_c^+ D^-) \nonumber\\
  &=& 4\Gamma(P_{\Sigma^0}\to \Xi_c^0 \overline D^0) \nonumber\\
  &=& 12\Gamma(P_{\Lambda}\to \Xi_c^+ D^-) \nonumber\\
  &=& 12\Gamma(P_{\Lambda}\to \Xi_c^0 \overline D^0) \,,\nn
\end{eqnarray}
\begin{eqnarray}
  3\Gamma(P_{\Sigma^-}\to \Sigma_{c}^{0}  D^-_s)
  &=& 3\Gamma(P_{n}\to \Sigma_{c}^{0} \overline D^0) \nonumber\\
  &=& 3\Gamma(P_{\Sigma^+}\to \Sigma_{c}^{++}  D^-_s) \nonumber\\
  &=& 3\Gamma(P_{\Sigma^0}\to   \Sigma_{c}^{+}   D^-_s) \nonumber\\
  &=& 3\Gamma(P_{p}\to \Sigma_{c}^{++} D^-) \nonumber\\
  &=&4\Gamma(P_{\Lambda}\to \Xi_{c}^{\prime+} D^-)\nonumber\\
  &=& 4\Gamma(P_{\Lambda}\to \Xi_{c}^{\prime0} \overline D^0) \nonumber\\
  &=& 6\Gamma(P_{\Sigma^+}\to \Xi_{c}^{\prime+} \overline D^0)\nonumber\\
  &=& 6\Gamma(P_{p}\to \Sigma_{c}^{+} \overline D^0) \nonumber\\
  &=& 6\Gamma(P_{\Sigma^-}\to \Xi_{c}^{\prime0} D^-) \nonumber\\
  &=& 6\Gamma(P_{n}\to \Sigma_{c}^{+} D^-) \nonumber\\
  &=& 12\Gamma(P_{\Sigma^0}\to \Xi_{c}^{\prime0} \overline D^0) \nonumber\\
  &=& 12\Gamma(P_{\Sigma^0}\to \Xi_{c}^{\prime+} D^-) \,.
\end{eqnarray}

\begin{table*}
\setlength{\tabcolsep}{20pt}
\caption{The amplitudes for strong decays of pentaquark.}\label{tab:P8_4}\renewcommand{\arraystretch}{1.6}
\begin{tabular}{l c l c}\hline
			channel & amplitude & channel & amplitude \\\hline
$P_{\Lambda}\to   \Lambda^0  J/\psi $ & $ -d_1$ &
   $P_{\Sigma^-}\to   \Sigma^+  J/\psi $ & $ -d_1$\\
   $P_{\Sigma^0}\to   \Sigma^0  J/\psi $ & $ -d_1$ &
   $P_{\Sigma^+}\to   \Sigma^-  J/\psi $ & $ -d_1$\\
   $P_{p}\to   \Xi^-  J/\psi $ & $ -d_1$ &
   $P_{n}\to   \Xi^0  J/\psi $ & $ -d_1$\\		
   $P_{\Lambda}\to   \Lambda_c^+   D^-_s $ & $ -\sqrt{\frac{2}{3}} e_1$ & $P_{\Lambda}\to   \Xi_{c}^{\prime+}  D^- $ & $ -\frac{\sqrt{3}}{2}  e_2$\\
   $P_{\Lambda}\to   \Xi_c^+  D^- $ & $ -\frac{e_1}{\sqrt{6}}$ &	$P_{n}\to   \Sigma_{c}^{0}  \overline D^0 $ & $ -e_2$\\
   $P_{\Lambda}\to   \Xi_c^0  \overline D^0 $ & $ \frac{e_1}{\sqrt{6}}$ &	$P_{\Lambda}\to   \Xi_{c}^{\prime0}  \overline D^0 $ & $ \frac{\sqrt{3} e_2}{2}$\\
   $P_{\Sigma^-}\to   \Xi_c^0  D^- $ & $ e_1$&$P_{\Sigma^-}\to   \Sigma_{c}^{0}   D^-_s $ & $ e_2$\\
   $P_{\Sigma^0}\to   \Xi_c^+  D^- $ & $ \frac{e_1}{\sqrt{2}}$&	$P_{\Sigma^-}\to   \Xi_{c}^{\prime0}  D^- $ & $ -\frac{e_2}{\sqrt{2}}$\\
   $P_{\Sigma^0}\to   \Xi_c^0  \overline D^0 $ & $ \frac{e_1}{\sqrt{2}}$&	$P_{\Sigma^0}\to   \Sigma_{c}^{+}   D^-_s $ & $ e_2$\\
   $P_{\Sigma^+}\to   \Xi_c^+  \overline D^0 $ & $ -e_1$&	$P_{\Sigma^0}\to   \Xi_{c}^{\prime+}  D^- $ & $ -\frac{e_2}{2}$\\
   $P_{p}\to   \Lambda_c^+  \overline D^0 $ & $ e_1$&$P_{\Sigma^0}\to   \Xi_{c}^{\prime0}  \overline D^0 $ & $ -\frac{e_2}{2}$\\
   $P_{n}\to   \Lambda_c^+  D^- $ & $ e_1$&	$P_{\Sigma^+}\to   \Sigma_{c}^{++}   D^-_s $ & $ -e_2$\nonumber\\& &
   $P_{\Sigma^+}\to   \Xi_{c}^{\prime+}  \overline D^0 $ & $ \frac{e_2}{\sqrt{2}}$\nonumber\\&&
   $P_{p}\to   \Sigma_{c}^{++}  D^- $ & $ e_2$\nonumber\\&&
   $P_{p}\to   \Sigma_{c}^{+}  \overline D^0 $ & $ -\frac{e_2}{\sqrt{2}}$\nonumber\\&&
   $P_{n}\to   \Sigma_{c}^{+}  D^- $ & $ \frac{e_2}{\sqrt{2}}$\\
   \hline
\end{tabular}
\end{table*}

If we take $P_c(4312)$ as $P_p$ and $P_{cs}(4459)$ as $P_{\Lambda}$ in Eq.\,(\ref{multiPc}), the discovery cascade decay modes reported by LHCb Collaboration are
\begin{eqnarray}
  \Lambda_b^0 \to P_p \, K^- \to  J/\psi \, p \, K^- \,,\nn\\
  \Xi_b^- \to P_{\Lambda} \, K^- \to J/\psi \, \Lambda \, K^- \,.
\end{eqnarray}
Having the results in Sec.\,\ref{Production} and Sec.\,\ref{Strongdecay}, we can write down the cascade decay modes of b-baryon which might be useful for finding new pentaquark states. In addition, there are also cascade decay modes of B-meson with probability of being experimentally discovered. All of them are collected in Table \ref{TGoldenModes}.

\begin{table*}
\caption{Cascade decay modes of b-baryon and B-meson with potential to be experimentally discovered.}\label{TGoldenModes}\begin{tabular}{ccc||ccc}\hline\hline
   & Cascade & Channel & & Cascade & Channel \\\hline
   %1th line
   $\Lambda_b^0\to $ & $P_{n} + \overline K^0 \to$ & $n + J/\psi+ \overline K^0$ & $\overline B^0_s\to$ & $P_{n} + \overline n \to $ & $n + J/\psi + \overline n$ \\
   %2th line
   $\Lambda_b^0\to $ & $P_{n} + \overline K^0 \to$ & $\Lambda_c^+ + D^- + \overline K^0$ & $\overline B^0_s\to$ & $P_{n} + \overline n \to $ & $\Lambda_c^+ + D^- + \overline n$ \\
   %3th line
   $\Lambda_b^0\to $ & $P_{n} + \overline K^0 \to$ & $\Sigma_{c}^{0} + \overline D^0 + \overline K^0$ & $\overline B^0_s\to$ & $P_{n} + \overline n \to $ & $\Sigma_{c}^{0} + \overline D^0 + \overline n$ \\
   %4th line
   $\Xi_b^0\to $ & $P_{\Lambda} + \overline K^0 \to$ & $\Lambda^0 + J/\psi + \overline K^0$ & $\overline B^0\to$ & $P_{\Lambda} + \overline n \to$ & $\Lambda^0 + J/\psi + \overline n$ \\
   %5th line
   $\Xi_b^0\to $ & $P_{\Lambda} + \overline K^0 \to$ & $\Xi_{c}^{\prime+} + D^- + \overline K^0$ & $\overline B^0\to$ & $P_{\Lambda} + \overline n \to$ & $\Xi_{c}^{\prime+} + D^- + \overline n$ \\
   %6th line
   $\Xi_b^0\to $ & $P_{\Lambda} + \overline K^0 \to$ & $\Xi_{c}^{\prime0} + \overline D^0 + \overline K^0$ & $\overline B^0\to$ & $P_{\Lambda} + \overline n \to$ & $\Xi_{c}^{\prime0} + \overline D^0 + \overline n$ \\
   %7th line
   $\Sigma_{b}^{+}\to $ & $P_{p} + \overline K^0 \to$ & $p + J/\psi + \overline K^0$ & $\overline B^0\to$ & $P_{\Sigma^+} + \overline p \to$ & $\Sigma^+ + J/\psi + \overline p$ \\
   %8th line
   $\Sigma_{b}^{+}\to $ & $P_{p} + \overline K^0 \to$ & $\Lambda_c^+ + \overline D^0 + \overline K^0$ & $\overline B^0\to$ & $P_{\Sigma^+} + \overline p \to$ & $\Xi_c^+ + \overline D^0 + \overline p$ \\
   %9th line
   $\Sigma_{b}^{+}\to $ & $P_{p} + \overline K^0 \to$ & $\Sigma_{c}^{++} + D^- + \overline K^0$ & $\overline B^0\to$ & $P_{\Sigma^+} + \overline p \to$ & $\Sigma_{c}^{++} + D^-_s + \overline p$ \\
   %10th line
   $\Sigma_{b}^{-}\to$  & $P_{n} + K^- \to$ & $n + J/\psi + K^-$ & $\Lambda_b^0\to$ & $P_{n} + \pi^0 \to$ & $n + J/\psi + \pi^0$ \\
   %11th line
   $\Sigma_{b}^{-}\to$  & $P_{n} + K^- \to$ & $\Lambda_c^+ + D^- + K^-$ &  &  &  \\
   %12th line
   $\Sigma_{b}^{-}\to$  & $P_{n} + K^- \to$ & $\Sigma_{c}^{0} + \overline D^0 + K^-$ &  &  &  \\
\hline\hline
\end{tabular}
\end{table*}

One might notice the singly Cabibbo-suppressed decays of b-baryon are also presented in this table, since pentaquark has been identified through Cabibbo-suppressed process $\Lambda_b^0 \to P_c \, \pi^- \to J/\psi \, p \, \pi^-$ by LHCb \cite{LHCb:2016lve}. As the fact that most of the multiquark states ($X, Y, Z, P_c$) have been observed in the decays of B-meson and b-baryon, we anticipate some of the cascade modes in Table \ref{TGoldenModes} can be measured in the near future.

%%%%%%%%%%%%%%%%%%%%%%%
\section{Conclusions}
\label{conclusions}
%%%%%%%%%%%%%%%%%%%%%%%
In summary, we have studied the production of pentaquark through weak decays of b-baryon and B-meson under the flavor SU(3) symmetry. Amplitudes for various decay channels have been parametrized in terms of a few SU(3) irreducible amplitudes and a number of testable relations were provided. Then the strong decays of charmonium pentaquark have been discussed as well. Using these results, we have listed some cascade decay modes which are likely to be used for reconstructing pentaquark states in experiments.

Finally we stress that the charmonium pentaquark provides us a unique platform for understanding the nature of strong force. The pentaquark spectrum with hidden $c\bar c$ and three light quarks is very rich. Without a reliable theory for many-body quark interactions, we have to speculate the reason why the less massive pentaquarks, whose components are all light quarks, are not seen yet. The flavor SU(3) analysis has the potential to help to interpret the results of existed and future experimental searches of charmonium pentaquarks. Finding the new cascade decay modes of b-baryon and B-meson in Table \ref{TGoldenModes} could provide crucial evidence which would help resolve the current and longstanding puzzles in the exotic charmonium sector.

\section*{Acknowledgements}
We thank Prof. Wei Wang and Dr. Ya-Teng Zhang for valuable discussions. W.H.H and J.X. is supported in part by National Natural Science Foundation of China under Grant No. 12105247, the China Postdoctoral Science Foundation under Grant No. 2021M702957. Y.X is supported in part by National Natural Science Foundation of China under grant No. 12005294.

%%%%%%%%%%%%%%%%%%%%%%%%%%%%%%%%%%

\end{document}